\documentclass[11pt]{article}

\def\mytitle#1{\setcounter{equation}{0}
\setcounter{footnote}{0}
\begin{flushleft}\Large\textbf{#1}\end{flushleft}
\vspace{0.25cm}}
\def\myname#1{\leftline{{\large #1}}\vspace{-0.13cm}}
\def\myplace#1#2{\small\begin{flushleft}\textit{#1}\\
\texttt{#2}\end{flushleft}}

\def\myclassification#1{\small\noindent
Pacs numbers : 98.80.Cq, 98.80.-k
#1\vspace{0.5cm}}
\usepackage{graphicx}
\begin{document}
\mytitle{Is thermodynamics of the universe bounded by event horizon a Bekenstein system?}

\myname{Subenoy Chakraborty\footnote{schakraborty@math.jdvu.ac.in} }

\myplace{Department of Mathematics, Jadavpur University, Kolkata-32, India.}
{}

\begin{abstract}
In this brief communication, we have studied the validity of the first law of thermodynamics for the universe bounded by event horizon with two examples. The key point is the appropriate choice of the temperature on the event horizon. Finally, we have concluded that universe bounded by the event horizon may be a Bekenstein system and the Einstein's equations and the first law of thermodynamics on the event horizons are equivalent.
\end{abstract}
\myclassification{}

Since the end of the last century, there are series of observational evidences [1-3] which put standard cosmology into a big question mark. Either one has to introduce exotic matter(dark energy) having large negative pressure within the framework of Einstein gravity or one has to modify the gravity theory itself, so that observed present accelerating phase of the universe can be explained. On the other hand, due to this accelerated expansion the existence of event horizon is assured and it is relevant to examibe universe bounded by event horizon as a thermodynamical system. In this context,Wang et al [4] in 2006 investigated the laws of thermodynamics in an accelerating universe dominated by dark energy with a time dependent equation of state. They showed that both the first law and second law of thermodynamics are satisfied on the dynamical apparent horizon while thermodynamical laws break down on the cosmological event horizon. They were not able to rescue the thermodynamical laws by redefining any parameter. So they claimed that the cosmological event horizon is unphysical from the point of view of the laws of thermodynamics.

Further they pointed out that the apparent horizon is the largest surface whose interior can be treated as a Bekenstein system i.e. satisfies the Bekenstein's entropy/mass bound $S\leq 2 \pi RE $ and Bekenstein's entropy/area bound $S\leq \frac{A}{4}$. In case of event horizon, although the Bekenstein entropy/mass bound can be satisfied, the Bekenstein entropy/area bound is violated. So they concluded that the thermodynamic system outside the apparent horizon is no longer a Bekenstein system and the usual thermodynamic description breaks down.

In this short communication, we shall show the validity of the first law of thermodynamics by two examples with appropriate choice of temperature on the event horizon. In this connection we should mention that in recent past, generalized second law of thermodynamics has been shown [5] to be satisfied assuming the first law for the universe bounded by the event horizon as a thermodynamical system for various matter distribution and in different gravity theories.

Assuming the homogeneous and isotropic FRW model of the universe, the metric can locally be expressed in the form
\begin{equation}\label{1}
ds^{2}=h_{i j}\left(x^{i}\right)dx^{i} dx^{j}+R^{2} d\Omega_2^{2}
\end{equation}
where $i$, $j$ can take values $0$ and $1$.
The two dimensional metric
\begin{equation}\label{2}
d\gamma^{2}=h_{i j}\left(x^{i}\right)dx^{i} dx^{j}
\end{equation}
where
\begin{equation}\label{3}
h_{i j}=diag\left\{-1, \frac{a^2}{1-\kappa r^2}\right\}
\end{equation}
is referred to as the normal metric, with $x^{i}$ being associated co-ordinates($x^0=t, x^1=r$). R=ar is the area radius, considered as a scalar field in the normal two-dimensional space. Another relevant scalar quantity on this normal space is
\begin{equation}\label{4}
\chi(x)=h^{i j}\left(x\right)\partial_{i}R \partial_{j}R=1-\left(H^2+\frac{\kappa}{a^2}\right)R^2,
\end{equation}
with $\kappa=0, +1, -1$ for flat, closed and open model respectively.

Now the apparent horizon, a null surface is defined at the vanishing of the scalar, i.e.,

$~~~~~~~~~~~~~~~~~~~~~~~~~~~~~~~~~~~~~~~~~~~~~~~~~~~~\chi(x)=0$,

which gives
\begin{equation}\label{5}
R_A=\frac{1}{\sqrt{H^2+\frac{\kappa}{a^2}}}
\end{equation}
The surface gravity on the apparent horizon is defined as
\begin{equation}\label{6}
\kappa_{A}=-\frac{1}{2}\frac{\partial \chi}{\partial R}|_{R=R_{A}}=\frac{1}{R_A}
\end{equation}
and hence the usual Hawking temperature on the apparent horizon turns out to be
\begin{equation}\label{7}
T_A=\frac{|\kappa_A|}{2\pi}=\frac{1}{2\pi R_A}
\end{equation}

On the other hand, the event horizon is defined as
\begin{equation}\label{8}
R_E=a\int_{t}^{\infty}\frac{dt}{a},
\end{equation}
where the infinite integral converges if $a\sim t^\alpha$ with $\alpha>1$, i.e., the event horizon has relevance only in the accelerating phase. Usually in the literature, the Hawking temperature on the event horizon is defined similar to the apparent horizon(i.e., eq(7)) and one takes
\begin{equation}\label{9}
T_E=\frac{1}{2\pi R_E}
\end{equation}
This is also supported from the measurement of the temperature by a freely falling detector in a de-Sitter background(where both the horizons coincide) using quantum field theory[6].

In the present work, we shall define the temperature on the event horizon similar to the apparent horizon starting from the definition of surface gravity in eq(6), i.e., we define
\begin{equation}\label{10}
\kappa_{E}=-\frac{1}{2}\frac{\partial \chi}{\partial R}|_{R=R_{E}}=\frac{R_E}{{R_A}^2}
\end{equation}
So the Hawking temperature on the event horizon becomes
\begin{equation}\label{11}
T_E=\frac{|\kappa_E|}{2\pi}=\frac{R_E}{2\pi {R_A}^2}
\end{equation}

As flat FRW model is much relevant in the context of the Wilkinson Microwave Anistropy probe data [7] so we take $\kappa=0$ throughout the work. Also for flat model the two horizons are related by the relation
\begin{equation}\label{12}
R_A=\frac{1}{H}=R_H<R_E,
\end{equation}
so the Hawking temperature on the event horizon can now be written as
\begin{equation}\label{13}
T_E=\frac{H^2 R_E}{2\pi}
\end{equation}
Clearly from the inequality (12), we have
\begin{equation}\label{14}
T_A=\frac{H}{2\pi}<T_E
\end{equation}
Now we shall show the validity of the first law of thermodynamics for the following two dark energy models.\\\\

\textbf{A. Dark energy as a perfect fluid with constant equation of state:}

The Friedmann equations are
\begin{equation}\label{15}
H^2=\frac{8\pi G}{3}\rho~~~~,~~~~\dot{H}=-4\pi G\left(\rho+p\right)
\end{equation}
where $p=\omega\rho\left(\omega, a constant, -1<\omega<-\frac{1}{3}\right)$ is the equation of state of the dark energy(DE)-fluid having energy density $\rho$ and thermodynamic pressure p and they obey the conservation relation
\begin{equation}\label{16}
\dot{\rho}+3H\left(\rho+p\right)=0
\end{equation}
For this DE model of the fluid the scale factor grows with time as
\begin{equation}\label{17}
a\left(t\right)=t^\frac{1}{\alpha}~~~~~~~,~~~~~~~\alpha=\frac{3}{2}\left(1+\omega\right)~~,~~0<\alpha<1
\end{equation}
and the event horizon evolves linearly with time in the form
\begin{equation}\label{18}
R_E=\left(\frac{\alpha}{1-\alpha}\right){t}
\end{equation}
Now, the amount of energy flux across the horizon within the time interval dt as [1]
\begin{equation}\label{19}
-dE_H=4\pi R_h^2T_{a b}\kappa^{a} \kappa^{b} dt
\end{equation}
with $\kappa^{a}$ a null vector.
So for the event horizon we get,
\begin{equation}\label{20}
-dE=4\pi R_E^3H\rho \left(1+\omega \right)dt=\frac{\alpha dt}{G\left(1-\alpha\right)^{3}}
\end{equation}
Due to Bekenstein area-entropy relation we have
\begin{equation}\label{21}
S_E=\frac{\pi R_E^{2}}{G}
\end{equation}
So we have
\begin{equation}\label{22}
T_EdS_E=\frac{H^{2}R_E^{2}\left(HR_E-1\right)dt}{G}=\frac{\alpha dt}{G\left(1-\alpha\right)^{3}}
\end{equation}
Thus we have the first law: $-dE=T_EdS_E$ on the event horizon. It should be noted that to get the last equality in equation (20)
we have used the 1st Friedmann equation given in equation (15).\\\\

\textbf{B. Holographic DE model:}

We shall consider non interacting two fluid system- one in the form of holographic DE and the other component as dark matter. Here we choose a dark energy model which follows the holographic principle. Using effective quantum field theory with $R_E$ as the IR cut off, the energy density of the holographic DE is of the form [8]
\begin{equation}\label{23}
\rho_D=\frac{3c^2}{R_E^2},
\end{equation}
where c is a dimensionless free parameter. The Friedmann equations for the present two-fluid system are $\left(8\pi G=1=c\right)$
\begin{equation}\label{24}
H^2=\frac{1}{3}\left(\rho_m+\rho_D\right)~~~~and~~~~\dot{H}=-\frac{1}{2}\left(\rho_D+\rho_m+p_D\right)
\end{equation}
where $\rho_m$ is the energy density of the dark matter(dust) and $\rho_D$ and $p_D$ are the energy density and the thermodynamic pressure of the holographic DE with equation of state [9]
\begin{equation}\label{25}
\omega_D=-\frac{1}{3}-\frac{2\sqrt{\Omega_D}}{3c}.
\end{equation}
Here $\Omega_D=\frac{\rho_D}{3H^2}$ is the density parameter for the DE.
Now the flux of energy across the event horizon becomes
\begin{equation}\label{26}
-dE=4\pi R_E^{3}H\left(\rho_m+\rho_D+p_D\right)dt=\frac{3}{2}R_E^3H^3\left(1+\omega_D\Omega_D\right)dt,
\end{equation}
while
\begin{equation}\label{27}
T_EdS_E=\frac{3}{2}H^3R_E^3\left(1+\omega_D\right)dt.
\end{equation}
Thus we have $-dE\neq T_EdS_E$.
However, if we consider only the holographic DE fluid instead of two fluid system then $\Omega_D=1$ and the first law of thermodynamics is satisfied.

Thus we are able to show the validity of the first law of thermodynamics on the event horizon with the newly proposed temperature on the event horizon(given in eq (11)) for two perfect fluid models-one with constant equation of state and the other in the form of holographic dark energy. Moreover, it should be noted that in deriving the first law of thermodynamics we have to use the first Friedmann equation. So on the other way starting from the first law of thermodynamics on the event horizon one is able to derive the Einstein's field equations. Hence for the proposed temperature on the event horizon, the Einstein's equations and the first law of thermodynamics on the event horizon are equivalent at least for the two cited examples and universe bounded by the event horizon may be considered as a Bekenstein system. Therefore, we conclude that this modified temperature on the event horizon is the first step towards a general prescription for the validity of the first law of thermodynamics on the event horizon and hence this thermodynamical prescription with event horizon agrees(qualitatively) with observations. For future work, we shall attempt to formulate such a general description on the event horizon. \\\\

{\bf Acknowledgement :}
This work has been done during a visit to IUCAA, Pune, India. The author is thankful to IUCAA for their warm hospitality.\\\\

\end{document}